# Exploiting Correlation Among Data Items for Cache Replacement in Ad-hoc Networks

Noman Islam
Center for Research in Ubiquitous Computing
National University of Comp. and Emerging Sciences
Karachi, Pakistan
noman.islam@nu.edu.pk

Zubair A. Shaikh
Center for Research in Ubiquitous Computing
National University of Comp. and Emerging Sciences
Karachi, Pakistan
zubair.shaikh@nu.edu.pk

*Abstract*— Ad-hoc Networks are special types of communication networks that don't require any prior infrastructure to work. One of the crucial properties of such networks is their disconnected mode of operation. Due to the unpredictable nature of these networks, a node often has to work in offline mode with cached contents. Therefore, it is very vital in ad-hoc environment to determine the particular data to be stored in the cache and the position in the cache where that data has to be stored. In this paper, we have proposed a novel technique for cache replacement in Ad-hoc Network based on the mining of Association Rules. Using FP-Growth Association Rules Mining, the correlation among data items is computed and is then used as an important heuristics during cache replacement. We have evaluated the proposed technique using the simulator JIST/SWANS. A query routing algorithm for MANET is proposed and a simulation model is developed to analyze the proposed algorithm. The proposed cache replacement technique has been tested on different MANET configurations. The results reflect significant improvement in cache hit ratio with the proposed technique.

## I. INTRODUCTION

Ad-hoc Network has been a vibrant area of research for last few years. These networks differ from traditional networks mainly because of their infrastructure-less nature and dynamically changing topology. Hence, solutions proposed for research issues (like power, routing, security and data management etc.) for traditional networks are not sufficient for MANET. A great deal of research work is currently going on to address various issues of MANET.

Caching plays a vital role in data management in MANET due to its unique characteristics. It allows data items that are expected to be used in near-future to be saved at a location closer to the requester. Traditionally, most of the caching techniques exploit spatial or temporal properties of data items for cache replacement. These techniques include Least Recently Used (LRU) or Most Frequently Used (MRU) etc. However, these approaches are not very useful for MANET because of their inherent limitations and the dynamic nature of MANET. In this paper, we have proposed a novel cache replacement technique for MANET. It is based on the assumption that those data items that were accessed together frequently in past are correlated to each other and are anticipated to be used together in future as well. In order to determine this correlation, we have used association rules mining.

Let $D = \{d_1, d_2, ..., d_n\}$ be a set of $n$-item data set. Let $F_i = \{f_j : f_j \in D \wedge 1 \leq j \leq n\}$ represents a frequent item set i.e. all the elements of $F_i$ are correlated to each other and were used together in past. According to our assumption, all the elements of set $F_i$ are likely to be requested together in future. The association rule mining algorithm then produces the set $F$ of all frequent item sets:

$$F = \{F_1, F_2, ..., F_f\}$$

where, $f$ is the total number of frequent item sets.

Based on the results of association rules mining, a cache replacement algorithm has been proposed. The algorithm prioritizes the cache locations $C=\{c_1, c_2, ..., c_m\}$ on the basis of their contents and then finds out an appropriate slot $k$, $0 \leq k \leq m$ in the cache where a newly arrived data item $d_i$ has to be placed. A query routing algorithm has also been proposed and a simulation model is developed to test the proposed scheme. In the sections to follow, we will talk about the proposed technique in detail along with implementation discussions. Rest of the sections of this paper has been organized as follows. First, we will discuss the related work in the area of caching. Then we will discuss the proposed approach that is followed by implementation details. Finally, we conclude the paper with future work.

## II. RELATED WROK

There has been a significant amount of research going in the domain of Distributed and Mobile Database Systems to facilitate efficient Information Sharing. Replication schemes like[1] achieve this objective to some extent but are inherently limited by the node mobility and the pre-requisite of knowledge about operating environment. Caching is another way to improve data access. Caching schemes for the mobile and distributed systems comprises of: *Simple Caching*[2], *Hierarchical Caching*[3] and *Cooperative Caching*[4] etc. A Hierarchical Caching Scheme is based on routing cache requests along a hierarchy in a structured manner. Cooperative Caching that is based on cooperation of a number on nodes to perform caching has been an area of extensive research for last few years. [4] has proposed three schemes for cooperative caching based on routing support in MANET that includes: *CachePath* (caching the path of the data), *CacheData* (caching the data) and *HybridCache* (combination of cache path and cache data) scheme. Other cooperative caching strategies include *CoCa*[5] and *GroCoCa* [6] etc. [7] proposed a cache replacement policy called *Least Utility Value(LUV)* that is based on several factors like item access probability, item size, coherency, distance between user and the cache node etc. The author also proposed a cooperative caching algorithm – *Zone Cooperative* – that is based on LUV in which 1-hop neighbors nodes forms a zone. [8] presents an *On-Demand Cooperative Caching* architecture for multimedia objects using soft state signaling. Another related area of caching in which significant research has been carried out is semantic caching i.e. information is cached based on the semantics of the data being cached. [9] presents a semantic caching scheme based on Bayesian probability for image caching. [10] introduced group coherence as a new coherence criteria for MANET based on the unique features of MANET. To address issues of cache invalidation [11] has proposed a cache invalidation strategy. [12] proposed an approximation algorithm for cache placement in MANET that aims at minimizing the total access cost. The algorithm is greedy in nature and it iteratively selects data items with the maximum benefit and place that in to the cache.

III. AN ASSOCTION RULE-BASED CACHING REPLACMENT SCHEME FOR MANET

Fig 1 shows the proposed approach highlighting the main components along with interactions among them. The *Query Manager* is responsible for handling all the data item requests generated by users. It first searches for the requested items in the cache and if there is no cache hit, it uses the proposed query routing algorithm (shown in Fig 2) to search for the data items over the network. All the data requests that are issued in one transaction are saved in a *log database*. We define a *transaction* as a set of data item requests that are issued simultaneously in one session. A *session* is a time interval in which the pause between any two data requests is below a certain threshold γ. The *Association Rules Mining* component utilizes the log database records to discover correlation among the data items. The *mining component* runs periodically and computes the correlation among the data items based on log database's transactions. The data structure used for maintaining the log database is circular list. The computed results of data mining are stored in a list which is then used during cache replacement. In the next subsections, we are going to discuss the proposed query routing algorithm and cache replacement technique in detail.

A. *Query Routing Algorithm*

Fig 2 shows the Query Routing algorithm that we have used to test the proposed Cache Replacement algorithm in MANET. The query routing algorithm looks for the data item in local cache and if there is no cache hit, it propagates the request to its adjacent nodes. If a data item is found in cache, then its reply is generated. Any node receiving the QueryReply, first caches this reply in its local cache and then confirms if this reply is intended for itself or not. If the reply is intended for some other node, query reply is propagated ahead. To avoid network overhead, TTL value of all forwarded message is first checked to confirm if the message has expired or not.

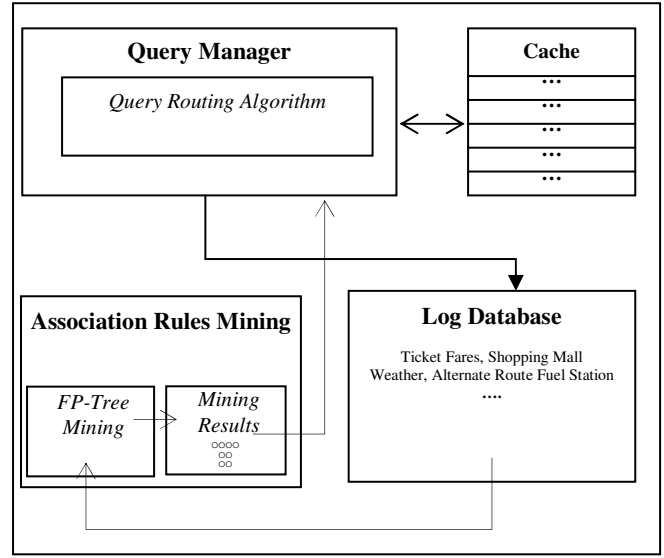

Figure 1. Proposed Caching Scheme Based on Association Rules Mining

```
[Query Routing Algorithm]
/*
C represents the Cache
Me represents the current host
*/

While (running) {
        Message m = read();
        if(m is QueryRequest) {
            if(lookup(m.dataitem,C) {
                Message res = m.createQueryReply(m);
                send(res);
            }else{
                forwardToAdjacentNodes(m);
            }
        }

        Else if(m is QueryResponse) {
            storeInCache(m.dataItem);

            if(m.destination <> Me) {
                forwardToAdjacentNodes(m);

            }
        }
}
```

Figure 2. Proposed Query Routing Algorithm

## B. Proposed Cache Replacement Algorithm

Let $C=\{c_1, c_2, c_3, \ldots, c_m\}$ represents the set of all cache entries with $c_i$ as individual locations. If we define an individual data items as $d_j$, then the cache replacement problem $CR(C, d_j)$ attempts to find out the optimum location $k$, $0 \leq k \leq m$, in the cache where item $d_i$ can be placed. The optimality criteria can be based on various factors like the frequency with which data items are accessed, the order in which data items are arrived in the cache etc. Let's now discuss the proposed cache replacement technique. Let $R_k$ represents the set of all data items that are correlated to $d_k$ based on the association rules mining algorithm. Then we define the residual set $S_k$ as:

$$S_k = C - R_k$$

where $S_k$ represents the residual set obtained after eliminating items related to $d_k$ from the set $S$. Then the particular cache location $j$ to be replaced with item $d_k$ can be computed as follows:

$$j = LRU(S_k)$$

where LRU returns the index of the least recently used item from the set $S_k$. Fig 3 shows the proposed cache replacement strategy in pseudo code format that tries to allocate a cache entry for a data item $d$. The algorithm first looks for an empty location in cache. If there is some space available in cache, the algorithm places the item $d$ in that free location. If there is no free space, the algorithm then calculate a victim cache location. To do so, the algorithm first constructs a residual set $S$ by eliminating those data items that are related to item $d$. If the residual set is non-empty, then the algorithm applies Least Recently Used (LRU) algorithm to determine cache location to be replaced with $d$. In some rare cases, the residual set might be empty. The algorithm applies LRU algorithm to all the cache contents in that particular case.

## IV. SIMULATION DETAILS AND RESULTS

The proposed caching strategy has been implemented in JIST/SWANS network simulation tool. Unless we explicitly mention, all the simulation parameters were kept to default values of JIST/SWANS. The simulation has been performed in a field of 500 × 500. The nodes were placed randomly. The routing algorithm we used is Ad hoc On Demand Distance Vector Routing (AODV) protocol. In order to test the proposed scheme, we need to generate some correlated traffic. Hence, a Correlated Data Generator (CDG) was developed in Java. The CDG module generates correlated data item requests periodically based on a correlation matrix $M$. Using Java Random Generator library, an n × n correlation matrix $M$ is generated in the following way:

$$M(i,j) = \begin{cases} 0 & r_{ij} < 0.5 \\ 1 & r_{ij} \geq 0.5 \end{cases}$$

where, $0 \leq i < n$ and $0 \leq j < n$ and $r_{ij} \in (0,1]$ is a random number generated for $M(i,j)$

Using this correlation matrix $M$, we generate the set of Candidate Data Item Requests $CDR$. To generate $CDR$, we first assign an index $i$ to all the data items. Then, a random data item $d$ is generated. The set $CDR$ is generated in the following way:

$$CDR = \{i \mid i = d \vee M(i,d) = 1\}$$

The set $CDR$ is generated based on whether the data item $i$ is related to data item $d$ based on the set $CDR$. Based on the set $CDR$, we then generated a request session $RS$ in the following way:

$$RS = \bigcup_{p_i < \eta} CDR(i)$$

where, $0 \leq i < n$ and $p_{ij} \in (0,1]$ is a random number generated for $RS(i,j)$ and $\eta$ is a threshold value that determines the likelihood of a data item request to be issued

In our case the value of $\eta$ is 0.8. There are a number of Association Rules Mining algorithm i.e. Aprioiri [13], FP-Tree[14]. We have selected FP-Tree algorithm because it is

```
[Cache Replacement Algorithm]
/*
    C represents cache memory locations
    item is the data item to be placed in cache
    S is the residual set
*/

public int placeItem(Cache C, Item d) {
    if(C.hasSpace() {
        C.add(item);
    }else{
        R = getRelated(d);
        S = C
        For i = 0 To R.size
        If S contains R(i)
            S.remove(R(i))
        End
    End

    int j;
    If size(S) <> 0
        j = getLRUSlot(S);
    Else
        j = getLRUSlot(C);

    C.setTouchTime(j, currentTime);
    C.set(j,d);
    }
}
```

Figure 3. Proposed Cache Replacement Algorithm

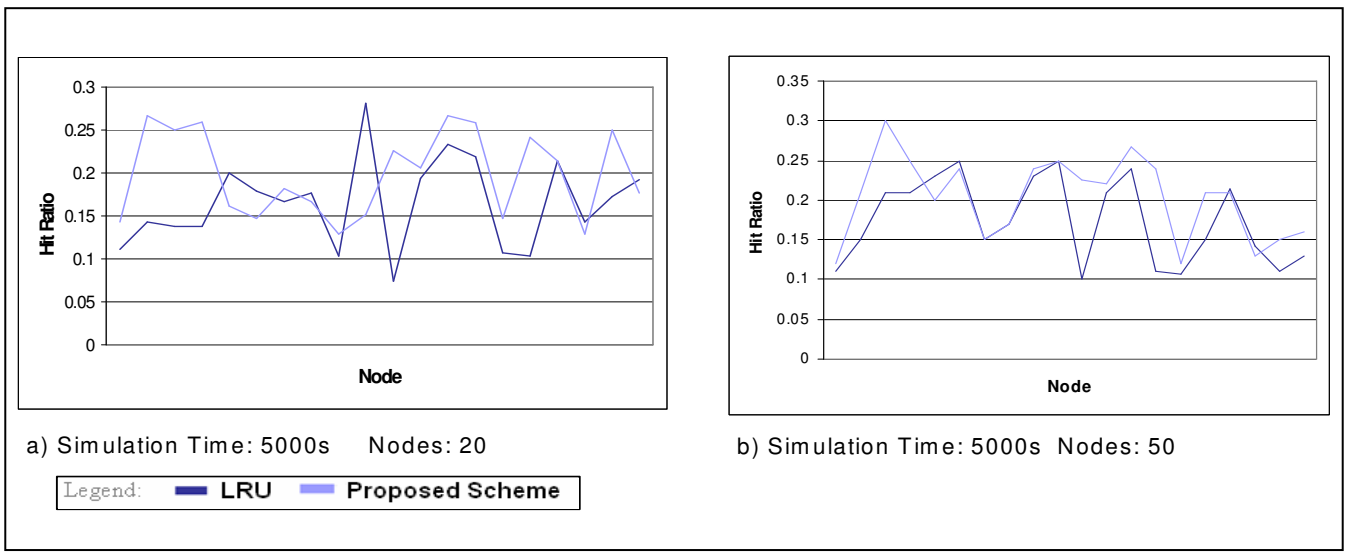

Figure 4. Results of Proposed Caching Technique

efficient in terms of processing time. The support value for FP-Tree mining algorithm is set to 80%. The capacity of the cache is set to 10 data items of fixed size. Fig 4 shows the results of the proposed scheme implemented on JIST/SWANS. The graph shows the hit ratio for each of the data item request. The graph compares the results of proposed cache replacement algorithm with LRU. Fig 4a shows the results for 10 nodes with simulation run for 5000s. As evident from graph, the proposed scheme performs reasonably well in contrast to LRU algorithm. Fig 4b shows the result with 50 nodes, that shows similar trends. In both of the results, there are some data item requests in which the LRU scheme performs better than proposed scheme. This can be because of the fact that the proposed scheme expects related data items to be used in future but in the actual case related items were not used together.

## V. CONCLUSION

In this paper, we have proposed a contemporary scheme for Cache Replacement in Mobile Ad hoc Network. The simulation results compared the suggested scheme with LRU cache replacement algorithm. The results show that the proposed scheme gives significant improvement in cache hit ratio as compared to LRU algorithm. However, the proposed scheme should be tested for various other parameters and configurations i.e. cache size, access time, number of nodes etc. In addition, the algorithm should also be tested when it is used in conjunction with other cache management algorithms i.e. cache consistency algorithms and cache update schemes etc. The proposed Cache Replacement scheme also incites several research areas. Since, the FP-Tree mining algorithm is not computationally good and its memory requirements are high, research should be pursued in extending existing FP-Tree mining algorithms such that it can work well on low-capability devices. A possible solution to this problem is designing approximation algorithms that give association results with reasonable accuracy and low computational cost.

ACKNOWLEDGMENT

This research work has been supported by Higher Education Commission, Pakistan and National University of Computer and Emerging Sciences, Karachi.

## VI. REFERENCES

1. Padmanabhan, P., et al., *A survey of data replication techniques for mobile ad hoc network databases.* The VLDB Journal, 2008. **17**(5): p. 1143-1164.
2. Yin, L. and G. Cao, *Supporting cooperative caching in ad hoc networks.* IEEE transactions on mobile computing, 2006. **5**(1): p. 77-89.
3. Bowman, C.M., et al., *The Harvest information discovery and access system.* Computer Networks and ISDN Systems, 1995. **28**(1-2): p. 119-125.
4. Cao, G., L. Yin, and C.R. Das, *Cooperative cache-based data access in ad hoc networks.* Computer, 2004: p. 32-39.
5. Chow, C.Y., H.V. Leong, and A. Chan. *Peer-to-peer cooperative caching in mobile environments.* 2004.
6. Chow, C.Y., H.V. Leong, and A.T.S. Chan. *Group-based cooperative cache management for mobile clients in a mobile environment.* 2004: IEEE Computer Society Washington, DC, USA.
7. Chand, N., R.C. Joshi, and M. Misra, *Cooperative caching in mobile ad hoc networks based on data utility.* Mobile Information Systems, 2007. **3**(1): p. 19-37.
8. Lau, W.H.O., M. Kumar, and S. Venkatesh. *A cooperative cache architecture in support of*

*caching multimedia objects in MANETs*. 2002: ACM New York, NY, USA.
9. Yang, B. and A.R. Hurson, *Semantic-Aware and QoS-Aware Image Caching in Ad Hoc Networks.* IEEE Transactions on Knowledge and Data Engineering, 2007. **19**(12): p. 1694-1707.
10. Alur, R. and M. Greenwald, *Coherency of Shared Memory in Ad-hoc Networks.* Technical Reports (CIS), 2001: p. 91.
11. Li, W., et al. *Cache Invalidation Strategies for Mobile Ad Hoc Networks*. 2007.
12. Tang, B., H. Gupta, and S.R. Das, *Benefit-based data caching in ad hoc networks.* IEEE Transactions on Mobile Computing, 2008. **7**(3): p. 289-304.
13. *Apriori algorithm*. [cited Aug, 09]; Available from: http://en.wikipedia.org/wiki/Apriori_algorithm.
14. Han, J., et al., *Mining frequent patterns without candidate generation: A frequent-pattern tree approach.* Data Mining and Knowledge Discovery, 2004. **8**(1): p. 53-87.